\documentclass[aps, prd, amsmath, floatfix, superscriptaddress, preprintnumbers, preprint]{revtex4-1}

\usepackage{amsmath, amssymb, amsfonts, mathtools, cases, slashed, bm, bbm}
\usepackage{array}
\usepackage{verbatim}
\usepackage{epsfig}
\usepackage{graphicx, graphics, color, epsfig, float}
\usepackage[colorlinks = true, citecolor = blue, linkcolor = blue, urlcolor = blue]{hyperref}
\usepackage[normalem]{ulem}
\usepackage[dvipsnames]{xcolor}
\usepackage{multirow}
\usepackage{graphicx}
\usepackage{dcolumn}
\usepackage{slashed}
\usepackage{siunitx}
\usepackage{xpatch}
\usepackage[mathlines]{lineno}
\usepackage{comment}
\usepackage{soul}
\usepackage{xcolor}
\usepackage{xfrac}
\usepackage{tabularx}
\usepackage{mathrsfs}
\usepackage{physics}
\usepackage{placeins}
\usepackage{dsfont}

\newcommand{\fref}[1]{Fig.~\ref{f.#1}}

\newcommand{\eref}[1]{Eq.~(\ref{e.#1})}

\newcommand{\tref}[1]{Table~\ref{t.#1}}

\begin{document}
\preprint{JLAB-THY-21-3368}

\title{Revisiting quark and gluon polarization in the proton at the EIC}

\author{Y. Zhou}
\affiliation{\mbox{Department of Physics, College of William and Mary, Williamsburg, Virginia 23187, USA}}
\author{C. Cocuzza}
\affiliation{\mbox{Department of Physics, SERC, Temple University, Philadelphia, Pennsylvania 19122, USA}}
\author{F. Delcarro}
\affiliation{Jefferson Lab,
	     Newport News, Virginia 23606, USA \\
        \vspace*{0.2cm}
        {\bf Jefferson Lab Angular Momentum (JAM) Collaboration
        \vspace*{0.2cm} }}
\author{W. Melnitchouk}
\affiliation{Jefferson Lab,
	     Newport News, Virginia 23606, USA \\
        \vspace*{0.2cm}
        {\bf Jefferson Lab Angular Momentum (JAM) Collaboration
        \vspace*{0.2cm} }}
\author{A. Metz}
\affiliation{\mbox{Department of Physics, SERC, Temple University, Philadelphia, Pennsylvania 19122, USA}}
\author{N. Sato}
\affiliation{Jefferson Lab,
	     Newport News, Virginia 23606, USA \\
        \vspace*{0.2cm}
        {\bf Jefferson Lab Angular Momentum (JAM) Collaboration
        \vspace*{0.2cm} }}

\begin{abstract}
We present a comprehensive impact study of future Electron-Ion Collider (EIC) data for parity-conserving and parity-violating polarization asymmetries on quark and gluon helicity distributions in the proton.
The study, which is based on the JAM Monte Carlo global QCD analysis framework, explores the role of the extrapolation uncertainty and SU(3) flavor symmetry constraints in the simulated double-spin asymmetry, $A_{LL}$, at small parton momentum fractions $x$ and its effect on the extracted parton polarizations.
We find that different assumptions about $A_{LL}$ extrapolations and SU(3) symmetry can have significant consequences for the integrated quark and gluon polarizations, for polarized proton, deuteron and $^3 \mathrm{He}$ beams.
For the parity-violating asymmetry, $A_{UL}$, we study the potential impact on the polarized strange quark distribution with different extrapolations of $A_{UL}$, finding the constraining power to be ultimately limited by the EIC machine luminosity.
\end{abstract}

\date{\today}
\maketitle

\section{Introduction}
\label{s:intro}

The quest to understand the spin decomposition of the proton into its primordial quark and gluon helicity and orbital angular momentum components has motivated considerable effort in the nuclear physics community for over 3 decades~\cite{Aidala:2012mv, Jimenez-Delgado:2013sma}.
The quark helicity contribution has been reasonably well constrained by polarized inclusive deep-inelastic scattering (DIS) data, from $x \approx 10^{-2}$ to $x \approx 0.5$, however, important questions still remain about its breakdown into individual quark flavors~\cite{Jimenez-Delgado:2013boa, Jimenez-Delgado:2014xza}.
The dominant $u$ quark helicity distribution, and to some extent the $d$ quark helicity, for instance, are well determined over a large range of parton momentum fractions~\cite{Sato:2016tuz}.
The strange quark helicity, on the other hand, is still rather elusive and difficult to constrain directly from data without additional theoretical assumptions~\cite{Ethier:2017zbq}.

The gluon distribution, which historically has been the most challenging to determine, is indirectly constrained by polarized DIS through scale evolution, but more directly from data on jet production in polarized $pp$ scattering at RHIC~\cite{Adamczyk:2014ozi}.
Unraveling the detailed spin structure of the nucleon has been a challenging task for several reasons.
The double-spin asymmetry $A_{LL}$ has provided the bulk of the constraints on the spin-dependent collinear parton distribution functions (PDFs).
However, in contrast to spin-averaged cross sections, the existing $A_{LL}$ data cover a considerably more limited range of kinematics, both in the four-momentum transfer squared, $Q^2$ ($Q^2 \lesssim 100$~GeV$^2$), and in the Bjorken-$x$ scaling variable ($x \gtrsim 0.01$).

\begin{figure}[t]
\centering
\includegraphics[width=0.8\textwidth]{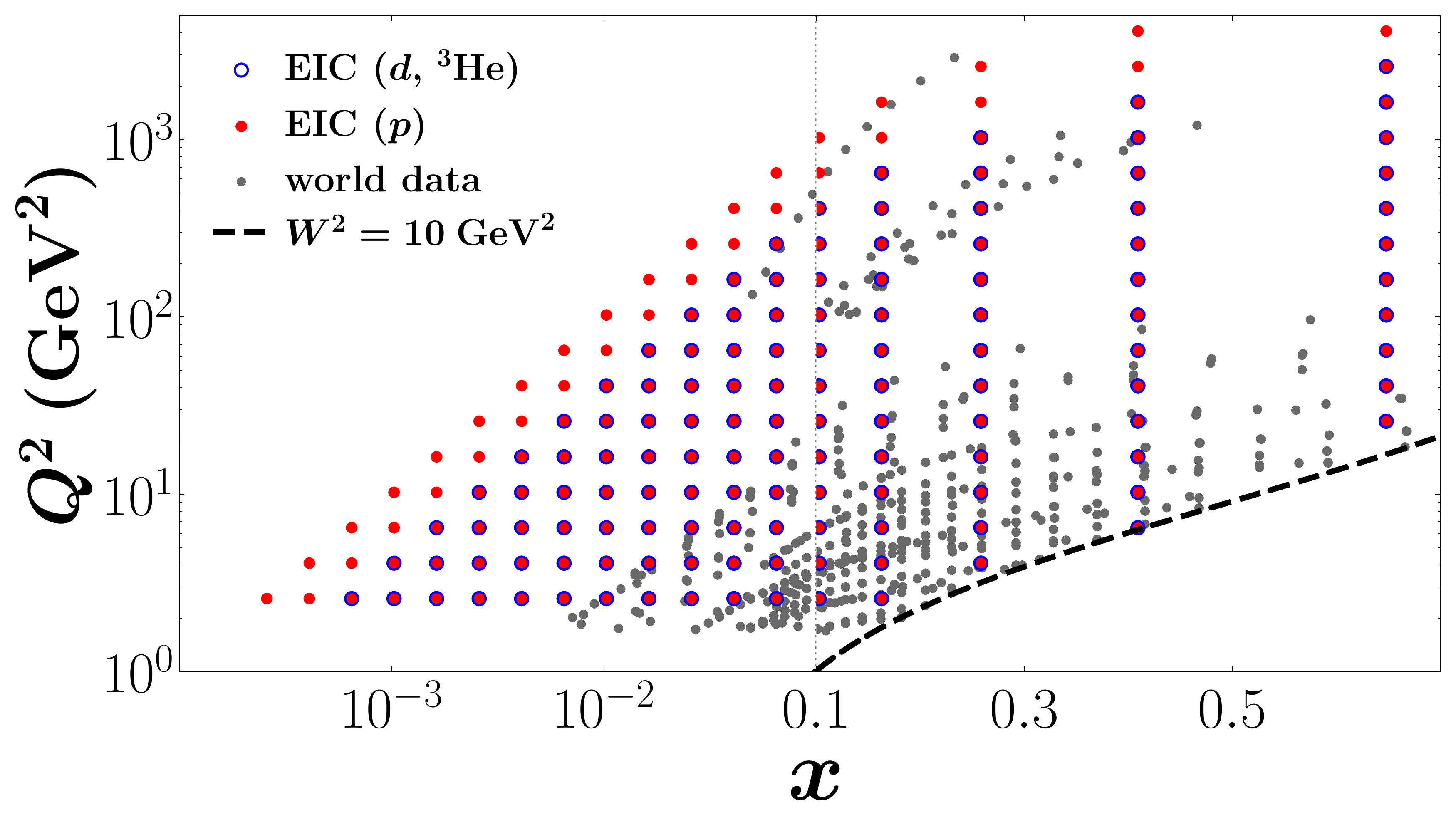}
\caption{Kinematic coverage of data sets used in this analysis, including spin-dependent fixed-target DIS (EMC, SMC, COMPASS, HERMES, SLAC), jet production in polarized $pp$ scattering (RHIC), and projected EIC data for polarized protons (red solid circles), deuterons and $^3$He (blue open circles). The variable $x$ here denotes Bjorken-$x$ for DIS and Feynman-$x$ for jet production in $pp$ collisions, while the scale $Q^2$ represents the four-momentum transfer squared for DIS and transverse momentum squared for jets. The dashed curve represents the boundary for the maximum $x$ attainable at fixed \mbox{$W^2 = M^2 + Q^2(1-x)/x = 10$~GeV$^2$}.}
\label{fig:kin}
\end{figure}

As the world's first polarized lepton-hadron (and lepton-nucleus) collider, the Electron-Ion Collider (EIC) will explore uncharted territory in spin physics~\cite{Accardi:2012qut}.
As illustrated in Fig.~\ref{fig:kin}, the EIC will extend the kinematic coverage in $x$ down to $x \approx 10^{-4}$, and in $Q^2$ up to $Q^2 \approx 10^3$~GeV$^2$.
Measurement of the polarization asymmetry $A_{LL}$ will access the $g_1$ structure function at unprecedented low values of $x$, and thus reduce uncertainties in spin-dependent PDFs at small parton momentum fractions.
Furthermore, the wider $Q^2$ coverage will allow scaling violations in the $g_1$ structure function to be determined more precisely, from which improved constraints can be derived on the spin-dependent gluon distribution.

Furthermore, access to polarized deuteron and $^3$He beams will allow separation of the helicity into individual quark flavors, and significantly reduce the uncertainties on the total helicity carried by quarks, $\Delta \Sigma$, compared to proton data alone, which are mostly sensitive to the $u$ quark polarization.
For the proton the available center of mass energies at the EIC are expected to be  $\sqrt{s} = 29, 45, 63$ and 141~GeV, while for deuteron and $^3$He the energies available would be $\sqrt{s} = 29, 66$ and 89~GeV, providing slightly smaller kinematic coverage than for the proton, as indicated in Fig.~\ref{fig:kin}.

With its high luminosity and hadron polarization compared to HERA \cite{Accardi:2012qut}, the EIC will also be able to access entirely new observables, such as the parity-violating asymmetry $A_{UL}$ involving unpolarized electrons and polarized hadrons~\cite{Anselmino:1993tc, Hobbs:2008mm}.
Dominated by the electromagnetic-weak interference term, this observable can provide new linear combinations of helicity PDFs that would enhance the flavor separation capabilities.
In particular, because of the electroweak couplings of the $Z$ boson to individual quark flavors, such data can provide a significant new constraint on the strange quark helicity distribution.

Previously there have been several dedicated analyses performed~\cite{Aschenauer:2015ata, Aschenauer:2020pdk} to assess the potential impact of future EIC data on DIS of polarized electrons from polarized hadrons.
In particular, Aschenauer {\it et al.}~\cite{Aschenauer:2020pdk} recently examined the impact of electron-proton $A_{LL}$ data at $\sqrt{s} = 44.7$~GeV within a full global QCD analysis.
Electron-proton and electron-$^3$He $A_{LL}$ pseudodata at $\sqrt{s}=141.4$~GeV and 115.2~GeV, respectively, were also studied using a reweighting method.
An earlier fit including projected electron-proton pseudodata at $\sqrt{s} = 77.5$, 122.7 and 141.4~GeV was performed in Ref.~\cite{Aschenauer:2015ata}.
Estimates of the impact of polarized semi-inclusive DIS data on flavor separation were made~\cite{Aschenauer:2020pdk} using reweighting of replicas~\cite{deFlorian:2019zkl} obtained by sampling the DSSV14 helicity PDFs~\cite{deFlorian:2014yva}, supplemented with the EIC proton pseudodata at $\sqrt{s}=44.7$~GeV.
Most recently, an impact study of projected EIC semi-inclusive DIS data for unpolarized PDFs and fragmentation functions was performed in Ref.~\cite{Aschenauer:2019kzf}.
A simulation study of neutral current structure functions in parity-violating DIS was made by Zhao {\it et al.}~\cite{Zhao:2016rfu}, using the DJANGOH event generator package~\cite{Charchula:1994kf} with the DSSV08 helicity PDFs~\cite{deFlorian:2008mr}.

In this paper we revisit the analysis of future EIC data on both inclusive parity-conserving and parity-violating polarization asymmetries and their impact on the quark and gluon helicity distributions in the proton.
Rather than relying on reweighting prescriptions, we account for the EIC pseudodata in a full global QCD analysis at next-to-leading order (NLO), using the JAM Monte Carlo framework~\cite{Sato:2016tuz, Ethier:2017zbq}, and include the most recent jet production data from unpolarized and polarized $pp$ collisions.
In addition to the proton and $^3$He data, we also study the impact of polarized electron-deuteron scattering data, which has not been considered in the previous analyses.
Most importantly, we critically examine the effect on the estimated PDF uncertainties of how the $g_1$ structure function, for protons, deuterons, and $^3$He, is extrapolated into the unmeasured low-$x$ region at $x \lesssim 0.01$.

We begin in Sec.~\ref{s.theory} by presenting the baseline PDF analysis on which the impact study will be built, including a summary of the observables used in this analysis, the PDF parametrizations employed, and an outline of the Bayesian inference methodology in the JAM Monte Carlo framework.
Section~\ref{s.simulation} discusses the simulation of the EIC observables and the specifics of the statistical and systematic uncertainty estimates.
The expected impact of the EIC proton, deuteron and $^3$He pseudodata for $A_{LL}$ and $A_{UL}$ on the $g_1$ structure functions and the quark singlet and gluon truncated moments is discussed in Sec.~\ref{s.impact}.
In particular, we carefully assess the role of low-$x$ extrapolation uncertainties and SU(3) flavor symmetry on the projected errors that can be attained with the new data.
Finally, in Sec.~\ref{s.conclusion} we summarize our conclusions and outline future work.

\section{Baseline analysis}
\label{s.theory}

In this section we present an overview of the theoretical framework used in the current analysis, including the EIC observables to be simulated, details of the PDF parametrizations employed, and the Bayesian inference methodology employed in the JAM Monte Carlo analysis.

\subsection{Longitudinal double-spin asymmetries}
\label{ss.theory_a_ll}

The double longitudinal spin asymmetry for the scattering of polarized electrons from polarized hadrons is defined as
\begin{equation}\label{eqn.a_ll}
A_{LL}
= \frac{\sigma^{\uparrow\Uparrow} - \sigma^{\downarrow\Uparrow}}
       {\sigma^{\downarrow\Uparrow} + \sigma^{\uparrow\Uparrow}}
= D\, \pqty{A_1 + \eta A_2},
\end{equation}
where \mbox{$\uparrow$ ($\downarrow$)} represents the spin of the lepton along (opposite to) the beam direction, and \mbox{$\Uparrow$ ($\Downarrow$)} represents the spin of the hadron along (opposite to) the beam direction.
The virtual photoproduction asymmetries $A_1$ and $A_2$ can be written in terms of ratios of the spin-dependent ($g_1$ and $g_2$) and spin-averaged ($F_1$ and $F_2$) structure functions,
\begin{align}
A_1 &= \frac{( g_1 - \gamma^2 g_2 )}{F_1},\ \ \ \ \ \
A_2  = \gamma \frac{( g_1 + g_2 )}{F_1},
\label{e.A}
\end{align}
and the kinematical variables in Eqs.~(\ref{eqn.a_ll}) and (\ref{e.A}) are defined as
\begin{eqnarray}
\begin{aligned}
D &=
\frac{y\, \pqty{2-y} \pqty{2 + \gamma^2 y}}
     {2 \pqty{1+\gamma^2} y^2 + \big[ 4 \pqty{1-y} - \gamma^2 y^2 \big] \pqty{1+R}}, 
\\
\eta &=
\gamma \frac{4 \pqty{1-y} - \gamma^2 y^2}{\pqty{2-y} \pqty{2 + \gamma^2 y}},
\end{aligned}
\end{eqnarray}
where $R$ is the ratio of longitudinal to transverse photoproduction cross sections and is given in terms of the spin-averaged structure functions,
\begin{align}
R &= \frac{ \pqty{1+\gamma^2} F_2 - 2x F_1 }{ 2x F_1 }.
\label{e.R}
\end{align}
The standard DIS variables $x$, $y$ and $\gamma^2$ are defined as
\begin{equation}
x = \frac{Q^2}{2 P \cdot q}, \qquad
y = \frac{P \cdot q}{P \cdot k} , \qquad
\gamma ^2 = \frac{4 M^2 x^2}{Q^2} ,
\end{equation}
where $P$, $k$ and $q$ are the four-momenta of the incident hadron (of mass $M$), incident electron, and exchanged virtual photon, respectively, with $Q^2 \equiv -q^2$.

At typical EIC kinematics (see Fig.~\ref{fig:kin}), one can take $M^2 \ll Q^2$, in which case $\gamma^2 \to 0$, $\eta \to 0$ and $R \approx F_2/2xF_1 - 1$, and the double polarization asymmetry simplifies to
\begin{equation}\label{e.A_LL}
A_{LL} = \frac{y\, (2-y)}{y^2 + 2 (1-y)(1+R)}\, \frac{g_1}{F_1}.
\end{equation}
At leading twist, the $g_1$ structure function, which in general is a function of $x$ and $Q^2$, can be expressed in terms of spin-dependent PDFs as
\begin{equation}\label{e.g_1_A_LL}
g_1(x,Q^2) = \frac12 \sum_q e_q^2\,
\Big( \big[ \Delta C_{1q} \otimes \Delta q^+ \big](x,Q^2)
    + 2 \big[ \Delta C_{1g} \otimes \Delta g \big](x,Q^2)
\Big),
\end{equation}
where $\Delta q^+ = \Delta q + \Delta \overline{q}$ is the sum of quark and antiquark helicity PDFs, $\Delta g$ is the gluon helicity PDF, and $\Delta C_{1q}$ and $\Delta C_{1g}$ are the perturbatively calculable hard scattering coefficients.
The symbol~``$\otimes$'' denotes the convolution integral
    $\big[\Delta C \otimes \Delta f\big](x)
    \equiv \int_x^1 (dz/z) \Delta C(z) \Delta f(x/z)$.

As indicated in \eref{g_1_A_LL}, the contribution from individual quark flavors $\Delta q^+$ is proportional to the square of their charges, so that at leading order the $g_1$ structure function of the proton is
    $g_1^p \approx (4 \Delta u^+ + \Delta d^+ + \Delta s^+)/18$.
Consequently, proton measurements mostly determine the $\Delta u^+$ PDF, and in order to constrain the $\Delta d^+$ and $\Delta s^+$ flavors one needs other hadrons, such as deuterons or $^3$He, or processes like semi-inclusive DIS, to provide additional combinations of the helicity PDFs (see Sec.~\ref{s.impact} below).

\subsection{Parity violating DIS}

A novel observable that can be studied at the EIC is the parity-violating asymmetry involving the scattering of unpolarized leptons from longitudinally polarized hadrons,
\begin{align}
A_{UL}
&= \frac{\sigma^{\Uparrow}-\sigma^{ \Downarrow}}
        {\sigma^{\Uparrow}+\sigma^{\Downarrow}},
\end{align}
where $\Uparrow$ ($\Downarrow$) denotes the spin of the hadron along (opposite to) the beam direction.
In this asymmetry the parity-conserving contributions from photon exchange and the vector-vector part of $Z$-boson exchange cancel exactly in the numerator, leaving the dominant contribution from the interference of photon and the axial-vector part of $Z$-boson exchange.
Neglecting the diagonal $Z$-exchange contributions and taking $M^2 \ll Q^2$, the parity-violating asymmetry can be written in terms of the spin-dependent interference $\gamma Z$ structure functions $g_{1,5}^{\gamma Z}$,
\begin{align}\label{e.A_PV}
A_{UL}
= \frac{G_F x Q^2}{2 \sqrt2 \pi \alpha}\,
  \bigg( \frac{g_A^e\, Y^- g_1^{\gamma Z} +\, g_V^e\, Y^+ g_5^{\gamma Z}}
       {x y^2 F_1 + (1-y) F_2}  \bigg),
\end{align}
where $G_F$ is the Fermi constant, $\alpha$ is the fine structure constant,
    $g_V^e = -\frac12 + 2 \sin^2\theta_W$
and
    $g_A^e = -\frac12$
are the vector and axial-vector couplings of the electron to the $Z$ boson, and $\theta_W$ is the weak mixing angle.
The kinematic factors in the numerator of (\ref{e.A_PV}) are given by
      $Y^\pm = 1 \pm (1-y)^2$.

At leading twist, the polarized $\gamma Z$ interference structure functions can be written in terms of the helicity PDFs as~\cite{Zhao:2016rfu},
\begin{eqnarray}
g_1^{\gamma Z}(x,Q^2)
&=& \sum_q e_q\, g_V^q
\Big( \big[ \Delta C_{1q} \otimes \Delta q^+ \big](x,Q^2)
    +2\big[ \Delta C_{1g} \otimes \Delta g \big](x,Q^2)
\Big),
\\
g_5^{\gamma Z}(x,Q^2)
&=& \sum_q e_q\, g_A^q\,
\big[ \Delta C_{5q} \otimes \Delta q^- \big](x,Q^2),
\end{eqnarray}
where $\Delta q^- = \Delta q - \Delta \overline{q}$ is the difference of quark and antiquark helicity PDFs, and the weak vector and axial-vector quark couplings are
    $g_V^{u,c} =   \frac12 - \frac43 \sin^2\theta_W$,
    $g_V^{d,s} = - \frac12 + \frac23 \sin^2\theta_W$,
and \mbox{$g_A^{u,c} = \frac12 = -g_A^{d,s}$}, respectively.
The contribution from the $g_5^{\gamma Z}$ structure function to $A_{UL}$ is suppressed by the small factor $g_V^e (\approx -4\%)$ and is generally negligible in the $x \lesssim 10^{-2}$ region.
The $g_1^{\gamma Z}$ structure function thus provides an independent linear combination of helicity PDFs, which, when combined with the electromagnetic $g_1$ structure function, can allow cleaner flavor separation.
For illustration, taking $\sin^2\theta_W \approx \frac14$ we can write $g_1^{\gamma Z}$ for the proton in the leading order approximation in terms of the quark singlet combination, 
\begin{align}\label{e:g1gZ}
g_1^{\gamma Z, p}(x,Q^2)
\approx \frac{1}{9} \Delta\Sigma(x,Q^2),
\end{align}
where $\Delta\Sigma = \sum_q \Delta q^+$.
In particular, this combination of helicity PDFs involves the $s$ quark on equal footing with the $u$ and $d$ quarks, making parity-violating DIS an exciting process for extracting the strange helicity distribution, to which existing polarized fixed target data have little sensitivity~\cite{Ethier:2017zbq}.
Furthermore, as a purely inclusive process, parity-violating DIS provides constraints on the flavor separation of PDFs, that are independent of SIDIS observables which rely on flavor tagging~\cite{Sato:2016wqj, Sato:2019yez}, and thus allow for new opportunities to test and validate the universality of PDFs.

\subsection{PDF parameterization}

Following previous JAM global QCD analyses~\cite{Sato:2016tuz, Ethier:2017zbq, Sato:2016wqj, Sato:2019yez}, we parameterize both the spin-averaged and spin-dependent PDFs at the input scale $\mu _0$ with a generic template function,
\begin{align}
{\rm T}(x, \mu_0; \bm{a})
= \frac{a_0}{{\cal N}}\, x^{a_1} (1-x)^{a_2} \big(1 + a_3 \sqrt{x} + a_4 x\big),
\label{e.template}
\end{align}
where $\bm{a} = \{ a_0, \cdots, a_4 \}$ denotes the set of shape parameters.
The normalization constant
    ${\cal N} = B(a_1 + n, a_2+1) + a_3 B(a_1 + n + \frac{1}{2}, a_2 + 1) + a_4 B(a_1 + n + 1, a_2+1)$,
where $B(x, y)$ represents the Euler beta function, is chosen to maximally decorrelate the overall normalization parameters from the shape parameters $\bm{a}$.
For the spin-averaged PDFs we take $n = 2$ so that $\mathcal{N}$ corresponds to the second moment used in the momentum sum rule, while for the spin-dependent PDFs we choose $n = 1$ so that $\mathcal{N}$ corresponds to the axial-vector charges.

For the spin-averaged PDFs, we take one shape for the valence $u_v = u - \bar{u}$ and $d_v = d - \bar{d}$ and gluon PDFs, while the sea quarks $\bar{u}$, $\bar{d}$, $s$ and $\bar{s}$ have independent shapes at high $x$ and an additional symmetric shape at low $x$.
For the spin-dependent PDFs, the parameterization is the same except that two shapes are taken for the gluon helicity for more flexibility and the sea quarks are taken to be symmetric at all values of $x$.  Additionally, all of the $a_3$ and $a_4$ parameters are fixed to zero for the spin-dependent PDFs.

Constraints on the normalization parameters $a_0$ for the valence helicity PDFs $\Delta u_v$ and $\Delta d_v$ are provided by the axial-vector charges,
\begin{subequations}
\label{e.su_2}
\begin{align}
\int_0^1 dx 
\big[ \Delta u^+(x,Q^2) - \Delta d^+(x,Q^2) \big] & = g_A,
\\
\int_0^1 dx
\big[ \Delta u^+(x,Q^2) + \Delta d^+(x,Q^2) - 2 \Delta s^+(x,Q^2) \big] & = a_8,
\end{align}
\end{subequations}%
where the triplet and octet axial-vector charges are obtained from neutron and hyperon $\beta$-decays~\cite{Jimenez-Delgado:2013boa},
\begin{subequations}
\label{e.charges}
\begin{eqnarray}
g_A &=& 1.269(3), \qquad\qquad\qquad {\rm [SU(2)]}
\label{e.su2}  \\
a_8 &=& 0.586(31), \qquad\qquad\qquad\!\!\! {\rm [SU(3)]}
\label{e.su3}
\end{eqnarray}
\end{subequations}%
using SU(2) and SU(3) flavor symmetry, respectively.
To assess the effect on the analysis of imposing SU(3) symmetry, we explore different scenarios of fitting only $g_A$ (the more flexible case) and fitting both $g_A$ and $a_8$ (the more constrained case) to their central values and uncertainties, as will be discussed further in Sec.~\ref{ss.baseline}.

To allow for greater speed of computation, Mellin space methods are used to solve the DGLAP evolution equations and the process dependent convolutions~\cite{Sato:2016tuz}. 
The renormalization group equations for the strong coupling and the evolution are solved numerically, with the QCD beta function evaluated at two loops with the boundary condition $\alpha_s(M_Z) = 0.118$.
The heavy quark mass thresholds for the evolution of the PDFs and $\alpha_s$ are chosen to be the PDG values $m_c = 1.28$~GeV and $m_b = 4.18$~GeV in the $\overline{\mathrm{MS}}$ scheme~\cite{Tanabashi:2018oca}.

\subsection{Bayesian inference}

For the execution of the global analysis including the projected EIC data, we use the fitting methodology developed by the JAM collaboration~\cite{Sato:2016tuz, Ethier:2017zbq, Sato:2016wqj, Sato:2019yez, Moffat:2021dji} based on Monte Carlo sampling of the parameter space with Bayesian inference.
Employing the multi-step strategy from Refs.~\cite{Sato:2019yez, Moffat:2021dji}, in the current analysis we first fit the spin-averaged PDFs using fixed target DIS data, after which HERA collider DIS data are included, followed by Drell-Yan and inclusive jet production data from hadronic collisions at the Tevatron and RHIC.
At this point the parameters of the spin-averaged PDFs are fixed, and spin-dependent PDFs are then fitted using first polarized DIS, then RHIC jet data from polarized $pp$ scattering, and finally the EIC pseudodata.
Of particular interest will be the impact on the helicity PDFs between the penultimate step (with all existing data) and with EIC pseudodata included.

From the ensemble of parameters $\{\bm{a}\}$ with dimension $N$ drawn from the posterior distribution, one can compute the expectation values and variances for any generic observable~${\cal O}$ (either a PDF at a given $x$ and $Q^2$, or a cross section computed from the PDFs),
\begin{subequations}
\label{e.EandVdef}
\begin{align}
{\rm E}[{\cal O}]
&= \frac{1}{N} \sum_k {\cal O}(\bm{a}_k),
\\
{\rm V}[{\cal O}]
&= \frac{1}{N} \sum_k \Big[ {\cal O}(\bm{a}_k)-{\rm E}[{\cal O}] \Big]^2,
\end{align}
\end{subequations}
where the variance gives the $1\sigma$ confidence interval for the observable ${\cal O}$.
The Bayesian ``master formulas'' (\ref{e.EandVdef}) provide the most robust determination of PDF uncertainties available within the global QCD analysis paradigm, without the need for introducing additional {\it ad hoc} prescriptions, such as tolerance factors, which are sometimes employed in single-fit analyses to account for tensions between data sets.

\section{EIC simulation}
\label{s.simulation}

In this section we present details of the simulation of the EIC pseudodata expected for measurements of the longitudinal double-spin asymmetry $A_{LL}$ and the parity-violating single-spin asymmetry $A_{UL}$, surveying various scenarios for center of mass energies, different hadrons, and integrated luminosities. 
To begin with, however, we first summarize the fit to existing unpolarized and polarized data used to determine the set of baseline spin-dependent PDFs which will be used to determine the impact of the future EIC data.

\subsection{Baseline PDFs}
\label{ss.baseline}

To simulate the impact of the EIC observables, we first need to obtain the unpolarized structure functions $F_1$ and $F_2$ which appear in the denominators of the polarization asymmetries in Eqs.~(\ref{e.A_LL}) and (\ref{e.A_PV}).
To this end, we perform a fit of the spin-averaged PDFs to fixed target inclusive DIS data from the BCDMS~\cite{Benvenuti:1989rh}, SLAC~\cite{Whitlow:1991uw}, and NMC~\cite{Arneodo:1996qe, Arneodo:1996kd} experiments, with cuts $W^2 = M^2 + Q^2 (1-x)/x > 10$~GeV$^2$ and $Q^2 > m_c^2$.
With the same cuts, we also include the reduced neutral current and charged current lepton-proton cross sections from the combined H1 and ZEUS analysis of HERA data~\cite{Abramowicz:2015mha}.
In addition, we include $pp$ and $pd$ Drell-Yan data from the Fermilab E866 experiment~\cite{Hawker:1998ty}, along with jet production data in $p\bar p$ collisions from D0~\cite{Abazov:2008ae} and CDF \cite{Abulencia:2007ez} at Fermilab, and in $pp$ collisions from STAR~\cite{Abelev:2006uq} at RHIC, with a transverse momentum cut $p_T > 10$~GeV.

The total $\chi^2$ value for the spin-averaged data is $\approx 3,665$ over 3,126 fitted data points, with a $\chi^2$ per degree of freedom $\chi_{\mathrm{dof}}^2 = 1.17$.
For the individual data sets we find $\chi_{\mathrm{dof}}^2$ values of 1.12 for the fixed target DIS data, 1.30 for HERA collider DIS data (with a combined value of 1.20 for all inclusive DIS data), 1.05 for Drell-Yan, and 0.97 for the $pp$ and $p\bar p$ inclusive jet data.
From the fitted spin-averaged PDFs we calculate the $F_1$ and $F_2$ structure functions, which are then kept fixed throughout the rest of the analysis of the spin-dependent data.

To obtain a baseline for the helicity PDFs, we perform a fit including fixed target polarized DIS data from EMC~\cite{Ashman:1989ig}, SMC~\cite{Adeva:1998vv, Adeva:1999pa}, COMPASS~\cite{Alekseev:2010hc, Alexakhin:2006oza, Adolph:2015saz},
SLAC~\cite{Baum:1983ha, Anthony:1996mw, Abe:1998wq, Abe:1997cx, Anthony:2000fn, Anthony:1999rm},
and \mbox{HERMES}~\cite{Ackerstaff:1997ws, Airapetian:2007mh},
with the same cuts on $Q^2$ and $W^2$ as the unpolarized DIS data~\cite{Sato:2016tuz, Ethier:2017zbq}.
In addition, we include jet production data from polarized $pp$ collisions from the STAR~\cite{Abelev:2006uq, Adamczyk:2012qj, Adamczyk:2014ozi, Adam:2019aml} and PHENIX~\cite{Adare:2010cc} collaborations, implementing a cut on the jet transverse momentum of $p_T > 10$~GeV.
%
The total $\chi^2$ for the spin-dependent data is $\approx 697$ over 696 fitted data points, for a $\chi_{\mathrm{dof}}^2 = 1.00$.
This includes a $\chi_{\mathrm{dof}}^2$ of 1.02 for the fixed target polarized proton and deuteron DIS data, and 0.78 for the jet data.
A more detailed discussion of the helicity PDFs with the DIS and jet constraints will be presented elsewhere~\cite{Zhou:2021}.

The baseline spin-averaged and spin-dependent PDFs determined from the global fit to the existing data are then used to simulate the impact of the observables at the EIC, assuming various scenarios for the theoretical assumptions.
In particular, we consider scenarios for $A_{LL}$ and $A_{UL}$ imposing only the SU(2) constraint in Eq.~(\ref{e.su2}) and also the SU(3) constraint in Eq.~(\ref{e.su3}), as well as assuming different behaviors for the low-$x$ extrapolation.

\subsection{Estimation of statistical and systematic uncertainties}
\label{ss.simulation}

The absolute statistical uncertainties for the polarization asymmetries are determined according to
\begin{align}
  \delta A \approx \frac{1}{\sqrt{\mathcal{L}\, \sigma_{\rm unp}}}, 
\label{e:statunc}
\end{align} 
where $\mathcal{L}$ is the estimated integrated luminosity for the specific process, and $\sigma_{\rm unp}$ is the unpolarized cross section.
This approximation is valid as long as the asymmetries are \mbox{$\ll 1$}.
Assuming that the cross sections can be considered constant in each $(x,Q^2)$ bin, the integrated unpolarized cross section can be written as
\begin{align}
\sigma_{\rm unp}
= \int_{\rm{bin}} dx\, dQ^2\,
\big( \sigma^{\uparrow\Uparrow} + \sigma^{\downarrow\Uparrow} \big)\,
\approx\, \Delta x \Delta Q^2\, \frac{8 \pi \alpha^2}{Q^2 s x^2}\,
\Big( x y F_1 + \frac{1-y}{y} F_2 \Big),
\end{align}
where $\Delta x$ and $\Delta Q^2$ are the intervals of the bins.

In the case of the $A_{LL}$ asymmetry, we assume a 2\% uncorrelated systematic uncertainty from the pion background, independent of the region of kinematics~\cite{ReneePrivate}.
Since the predictions for $A_{LL}$ are based on the extrapolation of existing measurements that are only available for \mbox{$x \gtrsim 0.01$}, we consider three possible scenarios, which we denote by ``low'',``mid'' and ``high'', to better assess the effect of extrapolation on the EIC pseudodata impact.
The high and low pseudodata sets are generated by shifting the values of $A_{LL}$ in the unmeasured region by $\pm 1\sigma$~CL, estimated from existing helicity PDF uncertainties, while the mid set is generated using the central predictions. 
For each data set, the uncertainties are shifted in the same way as the observables.

For the $A_{LL}$ asymmetry, we consider the center of mass energies $\sqrt{s} = 29$, 45, 63 and 141~GeV for a proton beam with an assumed integrated luminosity of 100~fb$^{-1}$, while for deuteron and $^3$He beams we include $\sqrt{s} = 29$, 66 and 89~GeV and assume 10~fb$^{-1}$ of integrated luminosity.
Projected $A_{LL}$ data and their uncertainties are shown in \fref{pseudo_data} for the mid scenario for proton, deuteron and $^3$He beams.
For the high (low) case, not shown in the figure, the small-$x$ region of the asymmetry will be shifted slightly upwards (downwards) by $\pm 1\sigma$~CL.
The systematic uncertainties follow the shape of the asymmetry, since they are estimated as a flat 2\% error.
The statistical uncertainties are similar for the high and low cases, as $\sigma_{\mathrm{unp}}$ in Eq.~(\ref{e:statunc}) is well-constrained down to $x \sim 10^{-4}$.

\begin{figure}[t]
\centering
\includegraphics[width=0.8\textwidth]{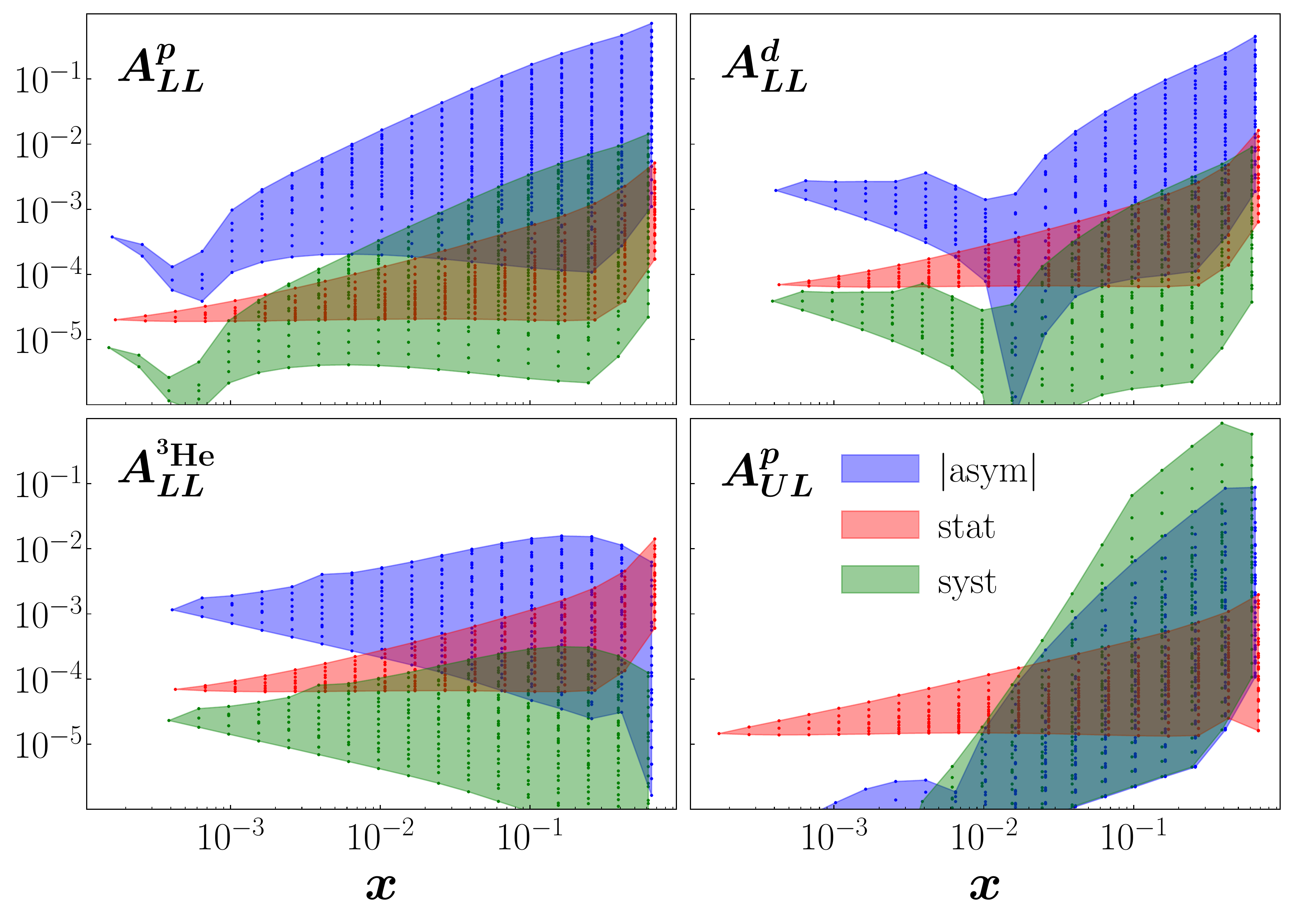}
\vspace*{-0.5cm}
\caption{Simulated absolute values of the longitudinal double-spin asymmetry $A_{LL}$ for proton, deuteron, and $^3$He beams and the parity-violating asymmetry $A_{UL}$ with a proton beam at the EIC (blue bands), using the ``mid'' predictions with both SU(2) and SU(3) assumptions from the baseline PDFs, along with estimated statistical (red bands) and uncorrelated systematic (green bands) uncertainties.}
\label{f.pseudo_data}
\end{figure}

\begin{table}[b]
\caption{Relative uncorrelated systematic uncertainties for $A_{UL}$ from the pion background, for various electron beam energies $E$ and pseudo-rapidity intervals $\Delta\eta$~\cite{ReneePrivate}.\\}
\begin{tabular}{c|c|c|c} \hline
$\Delta\eta$      & ~$E = 18$~GeV~ & ~$E = 10$~GeV~    & ~$E = 5$~GeV~ \\ \hline
~$(-3.5, -2.0)$~  & 0.02           & $10^{-3}$         & $10^{-5}$  \\
~$(-2.0, -1.0)$~  & 0.8            & 0.4               & 0.1        \\
$(-1.0, ~0.0)$~   & 1              & 8                 & 5          \\
~$(0.0, ~1.0)$    & 10             & 10                & 10         \\ \hline
\end{tabular}
\label{t.A_PV_syst}
\end{table}

For the parity-violating $A_{UL}$ asymmetry, we use the values given in \tref{A_PV_syst} for the predicted systematic uncertainties from the pion background, which are dependent on the electron beam energy $E$ and pseudo-rapidity $\eta = \ln(x \sqrt{s}/Q)$.
We consider the low, mid and high scenarios, as for the $A_{L L}$ asymmetry, and include only proton beam data at center of mass energies $\sqrt{s} = 29$, 45, 63 and 141~GeV, with an assumed integrated luminosity of 100~fb$^{-1}$.
The absolute values of the proton parity-violating asymmetry and the corresponding uncorrelated statistical and systematic errors are shown in the lower right panel of \fref{pseudo_data}. 
Note that estimates of correlated systematic uncertainties are not included in the current analysis.
Potential overall normalization errors will not affect the analysis, as the pseudodata are generated using the baseline PDFs described in Sec.~\ref{ss.baseline}.

\section{Impact of future EIC data}
\label{s.impact}

Having now established the theoretical framework and the set of baseline spin-averaged and spin-dependent PDFs, together with the estimated statistical and systematic uncertainties of the projected EIC data, in this section we present the results of the simulations including the $A_{LL}$ and $A_{UL}$ asymmetry pseudodata and their impact on the quark and gluon helicity distributions.
We consider a total of 6 scenarios for the $A_{LL}$ and $A_{UL}$ pseudodata, for each of the low, mid and high extrapolations below $x \sim 0.01$, and for both the more flexible SU(2) only case, fitting to $g_A$ as in Eq.~(\ref{e.su2}) and not enforcing SU(3), and the more restrictive case of fitting both SU(2) and SU(3) to $g_A$ and $a_8$ in Eqs.~(\ref{e.su2}) and (\ref{e.su3}), as summarized in \tref{scenarios}.

\begin{table}[h]
\caption{Summary of the 6 scenarios considered in this analysis for the baseline PDFs, with variations of the small-$x$ extrapolation (``low'', ``mid'', ``high'') and use of SU(2) and SU(3) constraints in Eqs.~(\ref{e.charges}) for the axial-vector charges. \\}
\begin{tabular}{c|ccc}
\hline
~scenario~ & ~extrapolation~ & ~SU(2)~ & ~SU(3)~ \\
\hline
1 & low  & \checkmark & \\
2 & mid  & \checkmark & \\
3 & high & \checkmark & \\
4 & low  & \checkmark & \checkmark \\
5 & mid  & \checkmark & \checkmark \\
6 & high & \checkmark & \checkmark \\
\hline
\end{tabular}
\label{t.scenarios}
\end{table}

\subsection{Constraints from $A_{LL}$ pseudodata}

\begin{figure}[t]
    \includegraphics[width=0.7\textwidth]{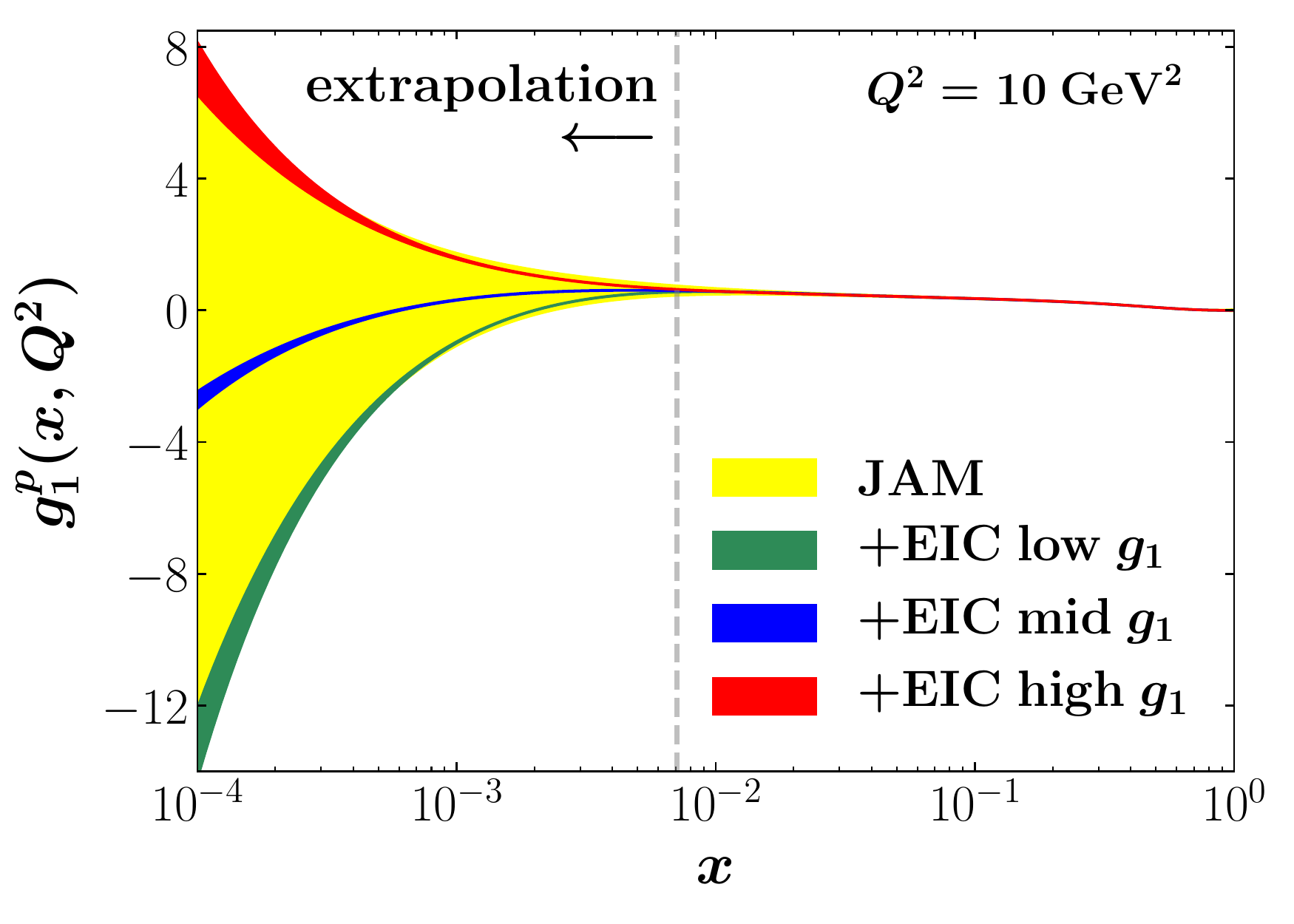}
    \vspace*{-0.5cm}
    \caption{Impact of projected $A^p_{LL}$ data at EIC kinematics on the proton $g_1^p$ structure function at $Q^2=10$~GeV$^2$, with the extrapolated baseline results (yellow band) compared with those including the EIC data for the $-1\sigma$ (``low'', green band), central (``mid'', blue band), and $+1\sigma$ (``high'', red band) uncertainties of $A_{LL}^p$, for the scenario of imposing both SU(2) and SU(3). The extrapolation region (indicated by the arrow) is to the left of the vertical dashed line at $x \approx 7 \times 10^{-3}$.}
    \label{f.g_1_p}
\end{figure}

The planned EIC experiments will extend measurements of $A_{LL}$ down to $x \approx 2 \times 10^{-4}$, which is almost 2 orders of magnitude smaller than the range of currently existing data.
In estimating the projected uncertainties on the data, a significant extrapolation of the $g_1$ structure function is therefore necessary into the unmeasured region. 
The extrapolation uncertainty is illustrated in Fig.~\ref{f.g_1_p} for the proton $g_1^p$ structure function at $Q^2=10$~GeV$^2$, extrapolated from the JAM baseline results as described in Sec.~\ref{s.simulation}.
The uncertainty on $g_1^p$ for $x \lesssim 10^{-3}$ is quite large, reflecting the absence of constraints from available measurements at low values of $x$.

The addition of EIC pseudodata leads to a dramatic reduction of the uncertainties, indicated by the colored bands in Fig.~\ref{f.g_1_p}, which represent extrapolations of $g_1^p$ according to the $-1\sigma$ (``low''), central (``mid''), and $+1\sigma$ (``high'') variations of $A_{LL}^p$.
The estimated uncertainties in this case are more comparable with the ones in the currently accessible $x$ region, suggesting the important constraints that can be anticipated from future EIC measurements.

\begin{figure}[t]
    \includegraphics[width=0.96\textwidth]{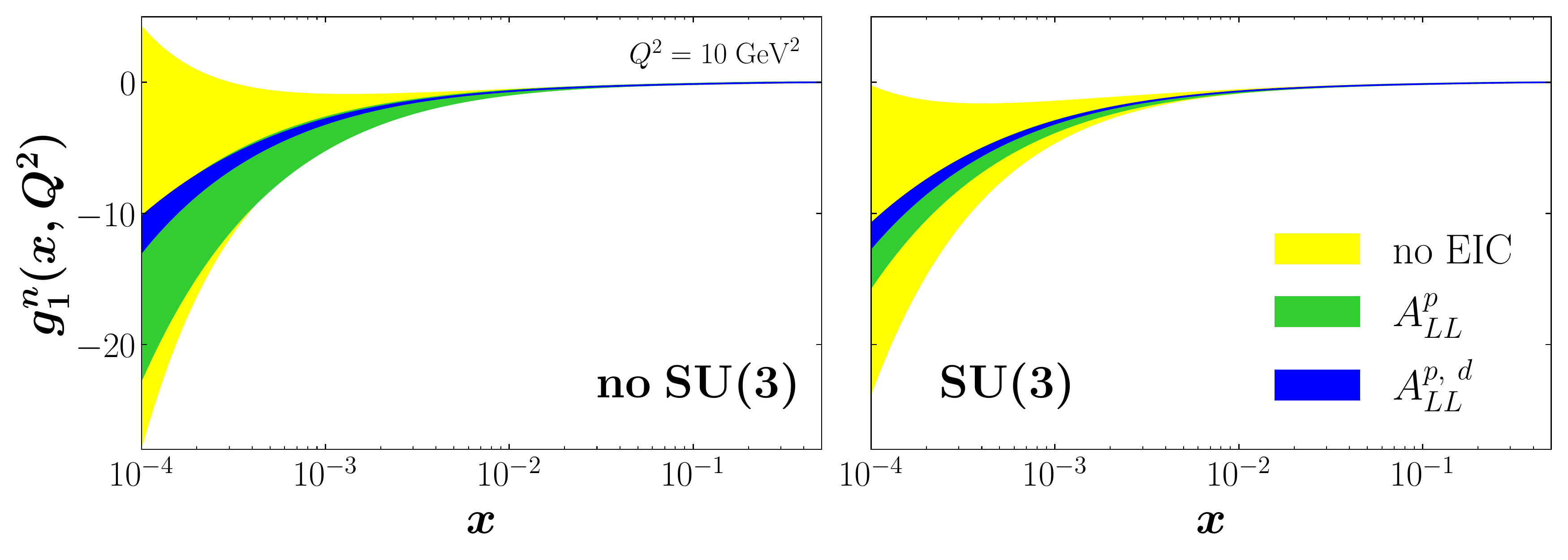}
    \vspace*{-0.5cm}
    \caption{Impact of projected proton $A^p_{LL}$ and deuteron $A^d_{LL}$ asymmetry data at EIC kinematics on the neutron $g_1^n$ structure function at $Q^2=10$~GeV$^2$ for the ``mid'' scenario. The extrapolated baseline results (yellow bands) are compared with those including EIC proton (green bands) and proton plus deuteron (blue bands) asymmetry pseudodata, for the case of not imposing SU(3) (left panel) and imposing SU(3) (right panel).}
    \label{f.A_LL_g1n_beams}
\end{figure}

The impact of the EIC $A_{LL}$ pseudodata on the neutron $g_1^n$ structure function is illustrated in Fig.~\ref{f.A_LL_g1n_beams} for the central (``mid'') scenario at $Q^2=10$~GeV$^2$.
From the figure one can see that while the proton pseudodata provide some constraints on $g_1^n$, further constraints are provided by the deuteron pseudodata, reducing the uncertainties by a factor of $2-4$ depending on whether SU(3) is imposed or not.
The same is observed if $^3$He pseudodata are used instead of deuteron (not shown in the figure).
This reduction of uncertainties on $g_1^n$ is correlated with a reduction of uncertainties on the $\Delta d$ PDF.

\begin{figure}[t]
    \includegraphics[width=0.6\textwidth]{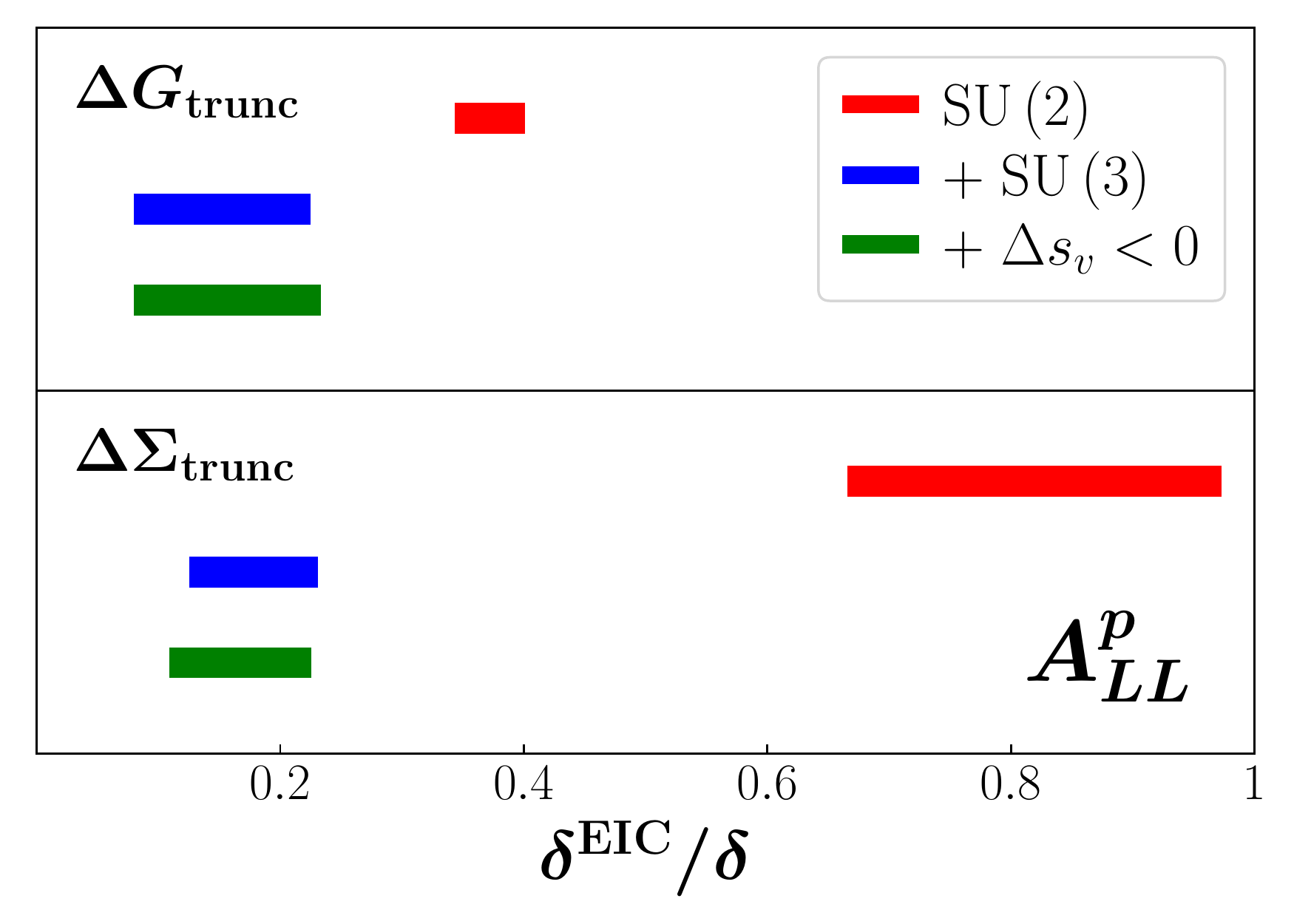}
    \vspace*{-0.5cm}
    \caption{Ratio of uncertainties $\delta^{\rm EIC}/\delta$ of the truncated moments of the gluon, $\Delta G_{\mathrm{trunc}}$ (upper panel), and quark singlet, $\Delta \Sigma_{\mathrm{trunc}}$ (lower panel), distributions with and without EIC data at $Q^2 = 10$~GeV$^2$, for scenarios of imposing only SU(2) (red bands), imposing SU(2) and SU(3) (blue bands), and in addition restricting solutions to ones with negative strangeness in valence region, $\Delta s_v < 0$ (green bands), using proton $A^p_{LL}$ EIC data. The ranges of the horizontal bands are obtained using uncertainties from the ``low'', ``mid'' and ``high'' scenarios in Fig.~\ref{f.g_1_p}.}
\label{f.A_LL_p_mom}
\end{figure}

To assess the impact of the EIC pseudodata on the spin carried by quarks and gluons in the proton, it is useful to consider truncated moments of the gluon and quark singlet helicity PDFs, defined as
\begin{eqnarray}
\label{e:deltaS}
\Delta G_{\mathrm{trunc}}(Q^2)
&\equiv& \int_{x_{\rm min}}^1 dx\, \Delta g(x,Q^2),
\\
\Delta \Sigma_{\mathrm{trunc}}(Q^2)
&\equiv& \int_{x_{\rm min}}^1 dx\, \sum_q \Delta q^+(x,Q^2),
\label{e:deltaG}
\end{eqnarray}
where the sum extends over the quark flavors $q = u$, $d$ and $s$, and in the present analysis we take $x_{\rm min} = 10^{-4}$.
Comparing the truncated moments and their uncertainties from the fits including the EIC proton $A^p_{LL}$ pseudodata and those from the baseline set of PDFs, in Fig.~\ref{f.A_LL_p_mom} we show the ratio of uncertainties $\delta^{\rm EIC}/\delta$ for both the gluon $\Delta G_{\mathrm{trunc}}$ and quark singlet $\Delta \Sigma_{\mathrm{trunc}}$ moment for all the scenarios listed in Table~\ref{t.scenarios}.

In the most general scenario, where only SU(2) symmetry is imposed via Eq.~(\ref{e.su2}), the impact of the $A^p_{LL}$ EIC pseudodata on $\Delta G_{\mathrm{trunc}}$ is an $\approx 60\%$ reduction of the uncertainty relative to the baseline fit uncertainty.
For the quark singlet moment $\Delta \Sigma_{\mathrm{trunc}}$, on the other hand, there is a much smaller, $\lesssim 30\%$, reduction in the uncertainty, which is effectively consistent with no reduction.
The ranges of the horizontal bands in Fig.~\ref{f.A_LL_p_mom} are obtained by considering the uncertainties from each of the ``low'', ``mid'' and ``high'' $A_{LL}$ scenarios in Fig.~\ref{f.g_1_p}.

The impact of the EIC pseudodata can increase when additional assumptions are made in the analysis.
In particular, by imposing SU(3) symmetry via Eq.~(\ref{e.su3}) the reduction of uncertainties on $\Delta G_{\mathrm{trunc}}$ is enhanced from $\approx 60\%$ to as high as $80\%-90\%$, with an even more dramatic improvement for the quark singlet moment.
The reduction of the latter can be understood from the fact that without the SU(3) constraint, both the $\Delta d^+$ and $\Delta s^+$ flavors are less well determined, and therefore contribute more to the uncertainty of $\Delta \Sigma_{\mathrm{trunc}}$.
The gluon distribution $\Delta g$, and the corresponding truncated moment $\Delta G_{\mathrm{trunc}}$, is less sensitive to SU(3) assumptions, hence the reduction in the uncertainty is more modest.

Note that our Monte Carlo analysis typically contains multiple solutions in parameter space, giving rise to fits with different shapes for poorly constrained distributions, which nevertheless yield essentially identical overall $\chi^2$ values.
This is especially relevant for the strange quark helicity PDF, $\Delta s^+$, which can be either positive or negative at intermediate $x$ values, $x \sim 0.1-0.3$, depending on whether the fit is constrained by semi-inclusive DIS data or not~\cite{Ethier:2017zbq}.
Typically, solutions with positive strange helicity in the valence region (``$\Delta s_v > 0$'') violate the SU(3) constraint, while the ones with negative strange helicity are more consistent with SU(3).
To avoid this violation, we consider in Fig.~\ref{f.A_LL_p_mom} also the scenario of restricting to negative polarized strangeness in the valence region (``$\Delta s_v < 0$'').
For the proton $A^p_{LL}$ pseudodata, however, the removal of the positive strange helicity solutions does not lead to any reduction in the uncertainty, since in this case the positive and negative $\Delta s_v$ have a very similar effect on $\Delta \Sigma_{\mathrm{trunc}}$ and its uncertainties.

\begin{figure}[t]
    \includegraphics[width=0.49\textwidth]{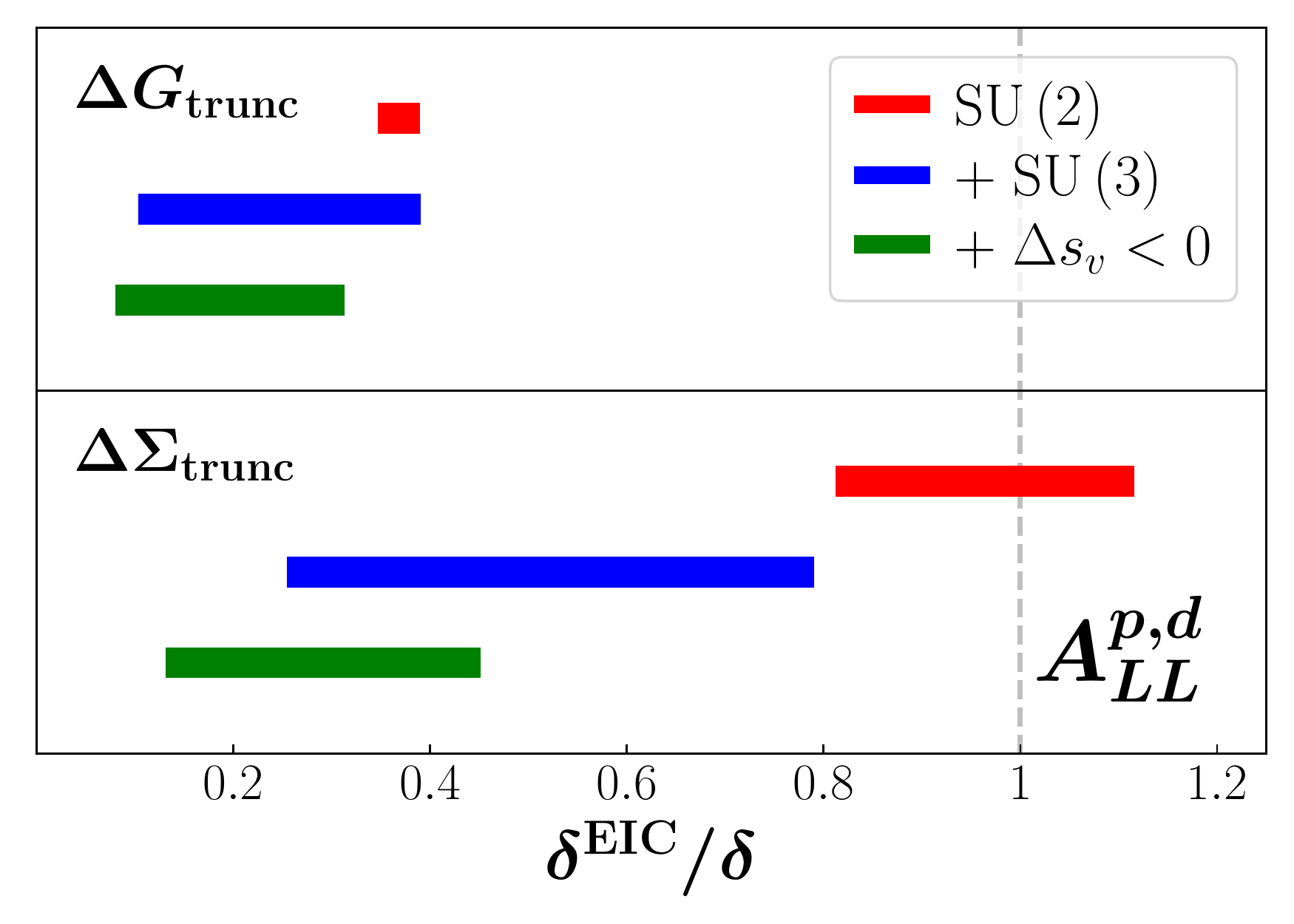}
    \includegraphics[width=0.49\textwidth]{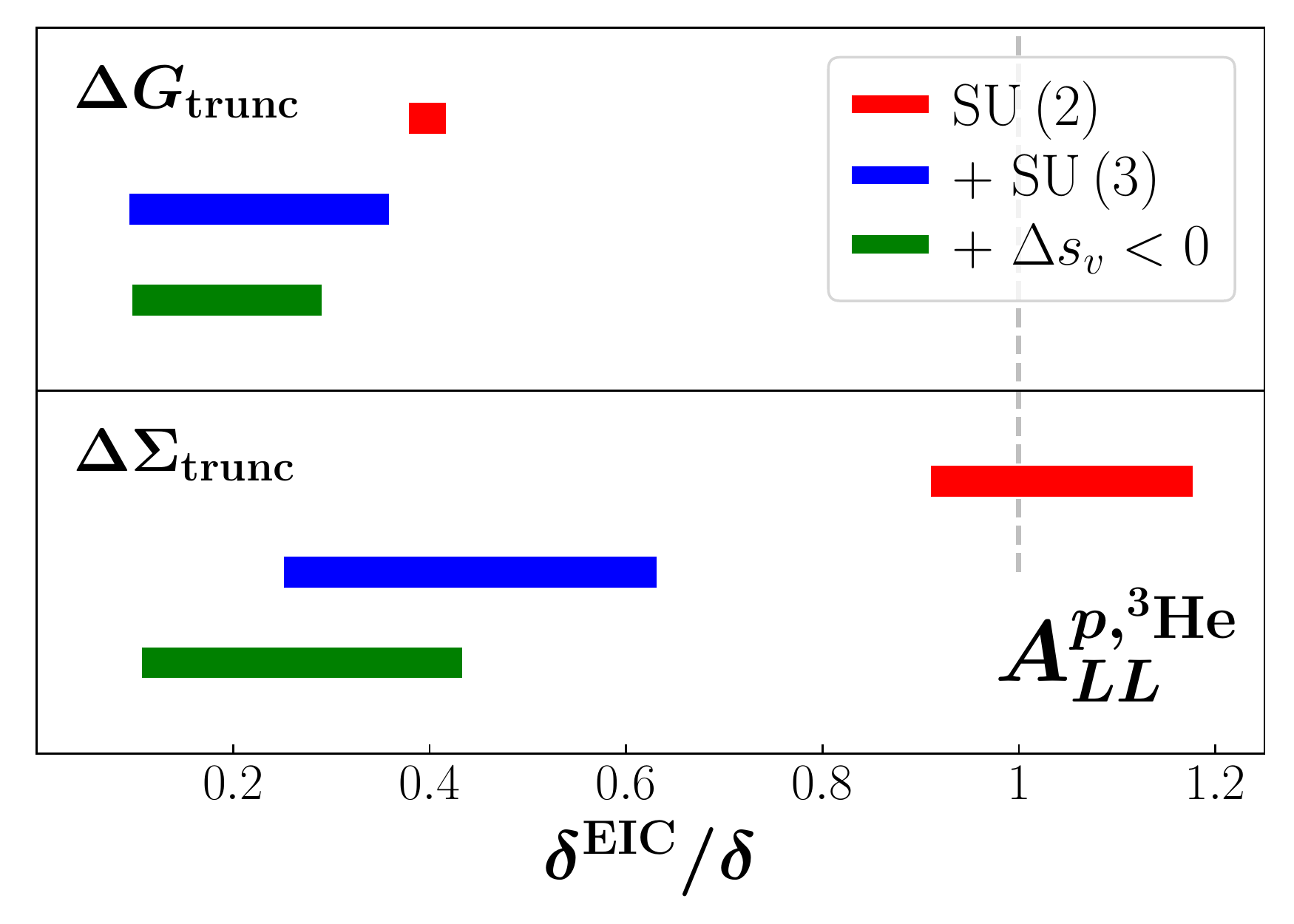}
\vspace*{-0.5cm}
    \caption{As for Fig.~\ref{f.A_LL_p_mom}, but considering the impact of proton $A^p_{LL}$ and deuteron $A^d_{LL}$ (left panel), proton $A^p_{LL}$ and helium $A^{^3\mathrm{He}}_{LL}$ (right panel) EIC pseudodata on the truncated gluon and quark singlet moments.}
\label{f.A_LL_n_mom}
\end{figure}

The effect of inclusion of $A_{LL}$ pseudodata for polarized deuteron and $^3$He beams is illustrated in Fig.~\ref{f.A_LL_n_mom}.
Here we observe an even clearer dependence of the impact for $\Delta \Sigma_{\mathrm{trunc}}$ on the theory assumptions made in the analysis.
When only the SU(2) constraint is imposed, no discernible impact on the quark helicity is observed.
After imposing SU(3), on the other hand, the impact on $\Delta \Sigma_{\mathrm{trunc}}$ ranges from $20\%-75\%$, depending on the low-$x$ extrapolation scenario.
If in addition the positive strange helicity solution is removed (due to its relatively large violation of SU(3)), the impact on $\Delta \Sigma_{\mathrm{trunc}}$ becomes more significant, with $60\%-90\%$ reduction in the uncertainty, and which is also less dependent on the extrapolation.

The impact on the gluon moment $\Delta G_{\mathrm{trunc}}$ from the inclusion of $A^d_{LL}$ or $A^{^3{\rm He}}_{LL}$ pseudodata is similar to the effect of using proton $A^p_{LL}$ data alone, with $\approx 60\%$ reduction in uncertainty for the combined $p + d$ or $p\, +\, ^3$He analyses.
This can be understood from the fact that the gluon contributes to the DIS asymmetry in essentially the same way for $p$, $d$ or $^3$He beams (appearing only at higher order in $\alpha_s$), so that addition of $d$ or $^3$He pseudodata does not improve the impact beyond what is already observed for $p$.
The further addition of SU(3) constraints or removal of $\Delta s_v > 0$ solutions does not significantly affect the impact on $\Delta G_{\mathrm{trunc}}$, since these constraints are largely indirect, with the overall reduction of uncertainties in the range $60\%-90\%$ in either the $p + d$ or $p\, +\, ^3$He scenarios.

We note, however, that both the SU(3) and $\Delta s_v < 0$ constraints are less justified than the constraint from SU(2), so that for the scenario that is least biased by theoretical input the impact of EIC $A_{LL}$ pseudodata is significant only for the gluon truncated moment $\Delta G_{\mathrm{trunc}}$ and is negligible for $\Delta \Sigma_{\mathrm{trunc}}$.

\subsection{Constraints from $A_{UL}$ pseudodata}

\begin{figure}[t]
\centering
\includegraphics[width=0.6\textwidth]{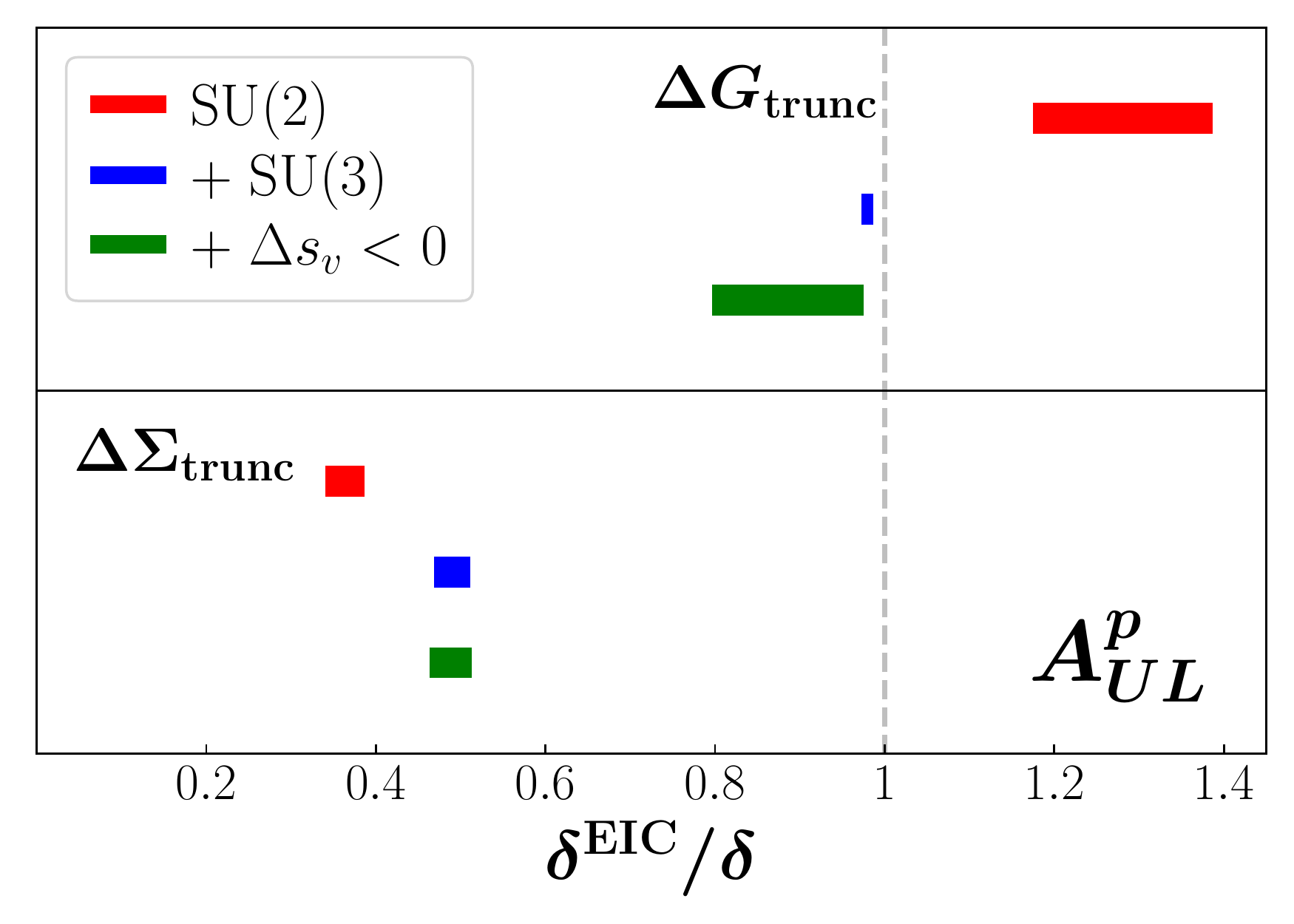}
    \vspace*{-0.5cm}
    \caption{As for Fig.~\ref{f.A_LL_p_mom}, but for the proton parity-violating $A^p_{UL}$ asymmetry.}
\label{f.AUL}
\end{figure}

The impact of the simulation described in Sec.~\ref{s.simulation} for the parity-violating proton single-spin asymmetry $A^p_{UL}$ is shown in Fig.~\ref{f.AUL}.
Interestingly, the situation here is somewhat inverted from that found for the $A_{LL}$ asymmetries in Figs.~\ref{f.A_LL_p_mom} and \ref{f.A_LL_n_mom}.
In particular, a strong impact is seen on the quark singlet truncated moment, with $\approx 50\% - 60\%$ reduction in the uncertainty for all three scenarios considered.
This result is in line with the expectation that the $g_1^{\gamma Z}$ structure function provides the dominant contribution to $A_{UL}$ [Eq.~(\ref{e.A_PV})] and weights the different quark flavor contributions approximately equally [Eq.~(\ref{e:g1gZ})].
Given that the baseline strange quark helicity distribution has weak constraints from existing data, the new $A_{UL}$ pseudodata are able to significantly improve the uncertainties on $\Delta s^+$, and thus on the quark singlet moment, $\Delta \Sigma_{\mathrm{trunc}}$.

On the other hand, no significant improvement is seen for the gluon truncated moment, regardless of the scenario considered.
Although the $g_1^{\gamma Z}$ interference structure function is as sensitive to the gluon distribution as is the electromagnetic $g_1$ structure function, the relative errors on the parity-violating $A_{UL}$ asymmetry are much larger than those on existing or projected $A_{LL}$ data (see Fig.~\ref{f.pseudo_data}).
It is therefore not surprising that the EIC $A_{UL}$ data are unable to provide significant new information on the gluon helicity distribution.
In fact, because of statistical fluctuations and the fact that the optimization of $\chi^2$ is performed on the observables rather than on the PDFs, it is possible in multidimensional fits such as the ones performed here to find an increase in PDF uncertainties in some regions of kinematics with inclusion of additional data~\cite{AbdulKhalek:2021gbh} (which does not occur at the observable level).

Finally, we note that in the EIC {\it Yellow Report}~\cite{AbdulKhalek:2021gbh}, the scenario of using SU(2) and SU(3) symmetry constraints from hyperon beta-decay was examined, and was found to have little impact on the quark singlet and gluon moments.
In the present, more robust analysis, the addition of a second shape for some of the helicity PDFs and the inclusion of a wider range of solutions for the gluon distribution give rise to an overall less well constrained baseline, and thus to a stronger impact on the quark singlet moment $\Delta \Sigma_{\mathrm{trunc}}$.

\section{Conclusions}
\label{s.conclusion}

With the plans for the construction of the next generation EIC facility now formally underway, we are on the threshold of an exciting new era of probing the structure of the nucleon with an unprecedented level of detail.
It is timely, therefore, to address the prospects of extracting physics from various observables planned for the EIC under specific projected experimental conditions.
In this paper we have revisited the extraction of the quark and gluon polarization from inclusive spin-dependent DIS measurements with polarized proton, deuteron and $^3$He beams, with a detailed impact study using the JAM Monte Carlo global QCD analysis framework~\cite{Sato:2016tuz, Ethier:2017zbq, Sato:2016wqj, Sato:2019yez, Moffat:2021dji}.

Expanding on the previous EIC impact studies in the literature~\cite{Aschenauer:2015ata, Aschenauer:2020pdk, deFlorian:2019zkl, Zhao:2016rfu, Aschenauer:2019kzf}, we have performed a global QCD analysis of existing polarized DIS and jet production data, which forms a baseline set of spin-dependent PDFs, together with EIC pseudodata on the longitudinal double-spin DIS asymmetry $A_{LL}$ and the parity-violating asymmetry $A_{UL}$.
We have explored for the first time the effects that different extrapolations into the unmeasured low-$x$ region can have on the degree of reduction of the uncertainties with the inclusion of the projected EIC data, along with the effects of assumptions about SU(2) and SU(3) flavor symmetry constraints on the axial-vector charges.

For the parity-conserving $A_{LL}$ asymmetry, we find that the impact of the EIC pseudodata on the quark singlet truncated moment $\Delta \Sigma_{\mathrm{trunc}}$ depends strongly on the SU(3) assumptions made, regardless of the use of $p$, $d$ or $^3$He beams.
For the most general case where no SU(3) symmetry is assumed, there is little reduction of the uncertainty.
The impact significantly increases, however, when SU(3) is imposed, and is also enhanced with the removal of solutions with positive strange helicity in the valence region that violate SU(3), especially with the inclusion of deuteron or $^3$He pseudodata.
The reduction of the uncertainty on the gluon truncated moment $\Delta G_{\mathrm{trunc}}$, on the other hand, is more robust at $\approx 60\% - 90\%$, and is less sensitive to the theoretical assumptions.

For the parity-violating $A_{UL}$ pseudodata, greater impact on $\Delta \Sigma_{\mathrm{trunc}}$ is found due to the unique combination of light quark flavors afforded by the $\gamma Z$ interference contributions.
The $\Delta G_{\mathrm{trunc}}$ moment, however, does not receive visible impact from the $A_{UL}$ pseudodata, mostly due to the relatively large projected uncertainties compared to the existing $A_{LL}$ asymmetry data.

The EIC facility will provide unprecedented access to the spin structure of the nucleon in previously unexplored regions of kinematics at low values of $x$.
Our analysis, performed within a robust Monte Carlo global QCD analysis of existing and projected data, should provide input for planning the future highest-impact observables to be measured at the EIC in optimal regions of kinematics.

\newpage
\section*{Acknowledgement}

We thank C.~Andres, J.~Qiu, and R.~Yoshida for helpful discussions.
This work is supported by the U.S. Department of Energy contract DE-AC05-06OR23177, under which Jefferson Science Associates, LLC, manages and operates Jefferson Lab. 
The work of C.C.~and A.M.~has been supported by the National Science Foundation under grant number PHY-1812359. The work of C.C. was also supported by Temple University.


\end{document}